\documentclass[twocolumn]{article}
\usepackage{graphicx}
\usepackage{amsmath}
\usepackage{amsfonts}
\usepackage{booktabs}
\usepackage{siunitx}

\usepackage[numbers]{natbib} 
\usepackage{hyperref}    
\usepackage{url}
\usepackage{braket}
\usepackage{pgfplots}
\usepackage{enumerate}
\usepackage{tikz}
\usetikzlibrary{math}
\usetikzlibrary{spy}
\tikzset{new spy style/.style={spy scope={%
			magnification=5,
			size=1.95cm, 
			connect spies,
			every spy on node/.style={
				rectangle,
				draw,
			},
			every spy in node/.style={
				draw,
				rectangle,
				fill=gray!10,
			}
		}
	}
}
\tikzmath
{
	function symlog(\x,\a){
		\yLarge = ((\x>\a) - (\x<-\a)) * (ln(max(abs(\x/\a),1)) + 1);
		\ySmall = (\x >= -\a) * (\x <= \a) * \x / \a ;
		return \yLarge + \ySmall ;
	};
	function symexp(\y,\a){
		\xLarge = ((\y>1) - (\y<-1)) * \a * exp(abs(\y) - 1) ;
		\xSmall = (\y>=-1) * (\y<=1) * \a * \y ;
		return \xLarge + \xSmall ;
	};
}

\title{Reverse quantum annealing assisted by forward annealing}
\author{Manpreet Singh Jattana\\Modular Supercomputing and Quantum Computing\\Kettenhofweg 139, Goethe University Frankfurt am Main}

\begin{document}

\twocolumn[
\begin{@twocolumnfalse}
	\maketitle
\abstract{Quantum annealers conventionally use forward annealing to generate heuristic solutions. Reverse annealing can potentially generate better solutions but necessitates an appropriate initial state. 
	Ways to find such states are generally unknown or highly problem dependent, offer limited success, and severely restrict the scope of reverse annealing. We use a general method that improves the overall solution quality and quantity by feeding reverse annealing with low-quality solutions obtained from forward annealing. An experimental demonstration of solving the graph coloring problem using the D-Wave quantum annealers shows that our method is able to convert invalid solutions obtained from forward annealing to at least one valid solution obtained after assisted reverse annealing for $57\%$ of $459$ random Erd\H{o}s--R\'enyi graphs. 
	Our method significantly outperforms random initial states, obtains more unique solutions on average, and widens the applicability of reverse annealing. Although the average number of valid solutions obtained drops exponentially with the problem size, a scaling analysis for the graph coloring problem shows that our method effectively extends the computational reach of conventional forward annealing using reverse annealing.
}

\quad
\end{@twocolumnfalse}
]

\section{Introduction}
Quantum annealers have made steady progress towards solving prototype real-world problems~\cite{Finnila1994,Harris2010,King2021,Ohzeki2020} along with gate-based quantum computers~\cite{cit22ekey,5088164,GYONGYOSI201951} and their hybrid applications~\cite{jattana2024quant,McClean2016,PhysRevApplied.19.024047,44,fphy.2022.907160,willsch2022hybridqu}. Quantum annealers are based on the principle of quantum annealing, which is based on the adiabatic theorem. Quantum annealing has been described as a metaheuristic for solving combinatorial optimization problems~\cite{022314,58.5355, ahaok}. 

Quantum annealing can be divided into two categories: forward annealing and reverse annealing (RA). The forward or conventional quantum annealing process starts in an initial state that is a uniform superposition of all possible classical states. The advantage of such an initial state is that it is an easy-to-prepare ground state of a known problem. The idea is to exploit the adiabatic theorem, whereby the amplitude of the transverse field is slowly decreased to zero, and the system remains in its instantaneous ground state at all times, thus yielding the solution to the optimization problem at the end of the process.

If we define the total Hamiltonian $H$ as

\begin{equation}
H = A(s)H_p + B(s)H_i, \label{eq1}
\end{equation}
then $H_i$ is the initial Hamiltonian and $H_p$ is the problem Hamiltonian. $A(s)$ and $B(s)$ describe the annealing schedule. In forward annealing, the function $A(s)$ is gradually increased from $0$ to $1$, i.e., $A(s): 0\to 1$ and $B(s): 1\to 0$, when $s: 0\to 1$ within a certain annealing time $t$. A majority of applications of quantum annealers involve this setup. However, several recent papers have studied or used reverse annealing~\cite{rannea1,022314,arxiv.2303.13748,Venturelli_2019,PhysRevA.100.052321, Chancellor_2017, PhysRevResearch.3.033006, pelofske2023simulating,pone.0244026, nsprob, imoto2023demonstration,Haba2022,kim2024xr,PhysRevApplied.17.054033,httpsRocutto} for various problems.

The process of reverse annealing, in contrast to forward annealing, does not start in an initial state corresponding to the superposition of all possible classical states. It starts precisely in only one of all possible classical states~\cite{rannea1, 022314}. The amplitude of the transverse field is thereafter first gradually increased to a certain point and then decreased. 

Thus, according to Equation~(\ref{eq1}), reverse annealing undergoes $A(s): 1 \to A(s^\prime) \to 1$, $B(s): 0\to B(s^\prime)\to 0$, when $s: 1\to s^\prime \to 1$, where $0<s^\prime<1$ is called the reverse distance. Reverse annealing has been shown to produce better results with a dependence on an appropriate choice of the initial state~\cite{rannea1, Chancellor_2017, arxiv.2303.13748, Venturelli_2019, PhysRevA.100.052321, PhysRevResearch.3.033006}. In contrast, an inappropriate choice of the initial state can be generally expected to lead to no improvement in the solution quality. Thus, such initial states play a decisive role in the effectiveness of reverse annealing.

Currently, it is not always straightforward to find the appropriate choice of initial state for a given problem. Previous works have used classical greedy algorithms~\cite{Venturelli_2019}, simulated annealing and approximate message passing algorithm~\cite{PhysRevResearch.3.033006}, and classical preprocessing or physical intuition~\cite{rannea1}. A general strategy is missing that can be put to use for any problem.

In contrast to forward annealing, which can be termed as a global search routine aimed at finding the global minimum, reverse annealing can be termed as a local search routine that refines existing solutions~\cite{pone.0244026}. The combination of both can potentially use the broad exploration capability of forward annealing and then perform local search refinement using reverse annealing~\cite{king2019quantumassisted}. However, it is unclear how to effectively combine these two in practice. 

In this work, we propose a method that uses forward annealing to provide good initial states for reverse annealing. To demonstrate the idea, we first use forward annealing to solve the graph coloring problem and increase the problem size until no valid solutions are sampled. We then show that reverse annealing can find valid solutions when invalid solutions obtained from forward annealing are used as the initial state. It also obtains more unique solutions. Additionally, using forward annealing to provide the initial state outperforms guessing it randomly. Our approach can be seen as improving the overall metaheuristic of quantum annealing with general applicability.

The rest of the paper is structured as follows. In Section~\ref{secgc}, we give a brief overview of the problem that we will solve in this work. In Section~\ref{secra}, we introduce the theory of RA and mention the relevant parameters at play in the process. In Section~\ref{secres}, we show several results of performing reverse annealing in four different cases, undertake a scaling analysis by increasing the problem sizes, compare our strategy against random initializations, and demonstrate the device independence of our idea. We discuss and conclude in Section~\ref{secconc}.

\section{Graph Coloring} \label{secgc}

Given an undirected graph $G = (V, E)$, where $V$ represents the set of vertices and $E$ represents the set of edges connecting these vertices, the graph coloring problem involves finding an order of coloring such that no two connected vertices receive the same color and all vertices are colored. The graph coloring problem has two sub-branches: the minimum-coloring problem and $k$-coloring problem. 

The minimum-coloring problem seeks to find the smallest number of colors $k$ to properly color $G$. This $k$ is also called the chromatic number 
of $G$. The $k$-coloring problem seeks to color $G$ using no more than $k$ colors. The former is NP-hard and the latter NP-complete for $k \geq 3 $~\cite{800119,574848}. 
In this work, we use the quantum annealer to solve the $k$-coloring problem. We create random graphs according to the Erdős--Rényi model using the \textit{networkx} package in Python~\cite{team2014networkx}. We set the value of $k$ equal to the number of colors needed by the `largest first' greedy algorithm.

The graph coloring problem is reformulated as a quadratic unconstrained binary optimization (QUBO) problem using a Python package~\cite{qubogen}, which is based on Ref.~\cite{glover} and can be solved on quantum annealers~\cite{dwave}. Some previous work has been undertaken that used quantum annealers: as an independent set sampler~\cite{kwok2020graph} and using a domain-wall encoding~\cite{9485068}. However, none of the previous works on reverse annealing known to the author~\cite{rannea1,022314,arxiv.2303.13748,Venturelli_2019,PhysRevA.100.052321, Chancellor_2017, PhysRevResearch.3.033006, pelofske2023simulating, nsprob, imoto2023demonstration} studied it for the graph coloring problem.

\section{Reverse Annealing}
\label{secra}

Reverse annealing can be categorized into different categories based on either the (a) Hamiltonian and its implementation or (b) the method used to produce the initial state for the first cycle of RA. In case of (a), RA is of three types: first, adiabatic RA, in which an additional term is added to $H$ given in Equation~(\ref{eq1}) and can be used to perform analytical treatment of a problem~\cite{PhysRevE.85.051112,PhysRevA.100.052321}; second, repetitive RA, in which no terms are added to Equation~(\ref{eq1}) and the initial state is always the same for each run~\cite{9259948}; third, iterated RA, in which no terms are added to Equation~(\ref{eq1}) but the output of a RA cycle is fed as input to sample the output from the next RA cycle. Iterated RA is an important tool because the progress of a previous cycle can be improved even further in a new cycle, in contrast to repetitive RA. If the initial state to iterated RA is a higher-order excited state, then it is possible to progress towards the ground state by reaching lower-order excited states through progressive cycles. Iterated RA is commonly used on D-Wave systems~\cite{citekeyira, Chancellor_2017}. We also use iterated RA for all our experiments. 

In case (b), RA is of at least four types: First, random input RA where an initial state is randomly generated; second, classically assisted RA, in which an algorithm on a classical computer produces an initial state; third, RA in which the initial state is motivated by physical or intuitive reasoning about the problem Hamiltonian; fourth, quantum-assisted RA, in which either the same or a different (type of) quantum device is used to produce the initial state. The novelty of our work lies in successfully combining this last category with iterated RA.

\begin{figure*}
\centering
\includegraphics[scale=1.1]{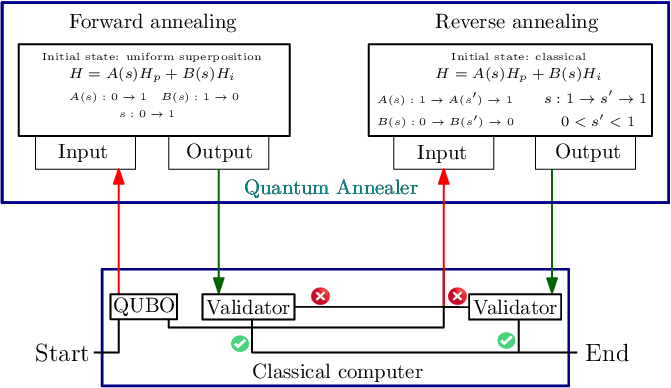}
\caption{{Working} 
of the forward-annealing-assisted reverse-annealing algorithm. The QUBO problem is first solved using forward annealing, and the algorithm halts if the output of the quantum annealer is a valid bitstring {(green check mark)}. Otherwise, the lowest energy bitstring is taken as input for reverse annealing, which can be made to run until a valid solution is obtained or a given sampling limit is reached.
\label{figmain}}
\end{figure*}

Since we use a quantum annealer to produce the initial state in the forward annealing setting, we call it forward-annealing-assisted RA. The algorithm is illustrated in Figure~\ref{figmain}. The problem graph is created on a classical computer and formulated into a QUBO problem. The quantum annealer then tries to solve the problem using forward annealing, and its output is fed to a validator, which checks for valid solutions among the bitstrings. Valid solutions are those that color our graphs in exactly $k$ colors without violating the constraints. If at least one valid solution is found among all the sampled bitstrings, then forward annealing has solved the problem and the algorithm ends. In contrast, if no valid solutions are found, one of the invalid solutions is input as the initial state for the first cycle of RA. One or more samples can be obtained through RA. The validator will check the output bitstring obtained from RA for validity. If an invalid solution is sampled, it is (within iterated RA) taken as an input for the next RA. The algorithm comes to an end if a valid solution is sampled. 

\subsection*{RA Parameters}\label{sec:oprev}
An important factor in the success of RA is the choice of parameters. These include the reverse distance, annealing schedule, and annealing path. We look at these parameters closely in this section.

In the following, we illustrate using a simple example of how choosing the reverse distance can be relevant and its relationship to different annealing schedules. To do so, we will look at the energy spectrum of $H$ from Equation~(\ref{eq1}) for one case of the graph coloring problem. Consider a graph with five vertices in the form of a one-dimensional chain without periodic boundaries. This graph can be colored using two colors. To proceed, we first convert the QUBO to its corresponding $H_p$. A given QUBO variable can be mapped to the corresponding Ising variable using

\begin{equation}
s_i = 1 - 2 x_i,
\end{equation}
where $s \in \set{+1,-1}$ is the $i$-th spin variable and $x\in \set{0,1}$ is the $i$-th binary variable. The Ising Hamiltonian can be written as~\cite{2014.00005, 9259934}

\begin{equation}
H_p = \sum_{i,j=1}^{N}Q_{ij}(\mathbf{1}-Z_i)(\mathbf{1}-Z_j), \label{eq3}
\end{equation}
where $Q$ is the QUBO matrix and $Z$ is the $\sigma^z$ Pauli operator. Python packages are available for converting QUBO to Ising and vice versa~\cite{dwave}.

\begin{figure}
\input{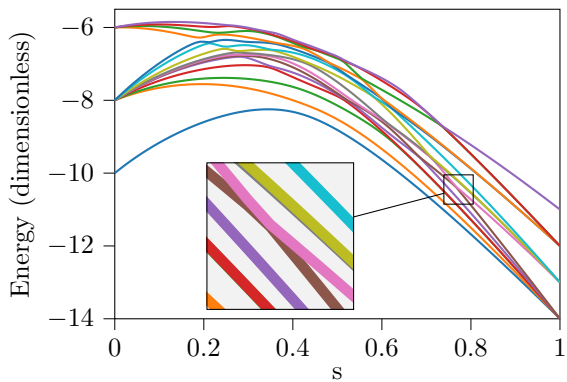}
\caption{Energy spectrum of a coloring problem with five vertices using a linear annealing schedule. Shown are the $15$ lowest-lying eigenvalues as a function of $s$, the annealing parameter. Inset: an anti-crossing between paths of ground and first excited states, shown in brown and pink, respectively. \label{fig:linear}}
\end{figure}

One must decide on an annealing scheme to study the energy spectrum in Equation~(\ref{eq1}). A linear scheme is one of the simplest. To use this scheme, we set
\begin{equation}
H = sH_p + (1-s) H_i, \label{eq2} 
\end{equation}
where $H_i = \sum_{j=1}^{N} \sigma^x_j$, and $s$ increases from $0$ to $1$ in one hundred equal steps. At each step, we reconstruct and exactly diagonalize $H$ and plot its $15$ lowest-lying eigenvalues. We include degenerate cases, if any. The spectrum is shown in Figure~\ref{fig:linear}.

The energy spectrum in Figure~\ref{fig:linear} reveals that the ground state is degenerate. This is understood intuitively by noting that, in general, there are multiple ways to color a graph. It is interesting to note that multiple paths lead to the ground state in our example. Several anti-crossings can be observed. We highlight one of them (in pink) in the inset magnified image, where one starts from the first excited state (at $s=1$), anneals back up to $s\approx 0.8$, and performs forward annealing; then, if the system shifts to another path (in brown), it leads to the ground state. Such a situation would imply that we started from an invalid solution initial state and obtained a valid solution final state.

\begin{figure}
\input{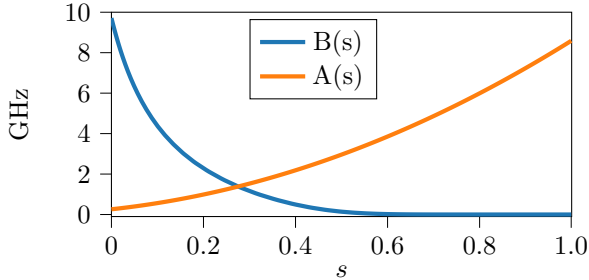}
\caption{Annealing schedule used by D-Wave quantum annealers. The parameter $s$ goes from $0$ to $1$ during forward annealing, whereas it goes from $1$ to some middle non-zero value and then back to $1$ for reverse annealing.\label{fig:sche}}
\end{figure}

\begin{figure}
\input{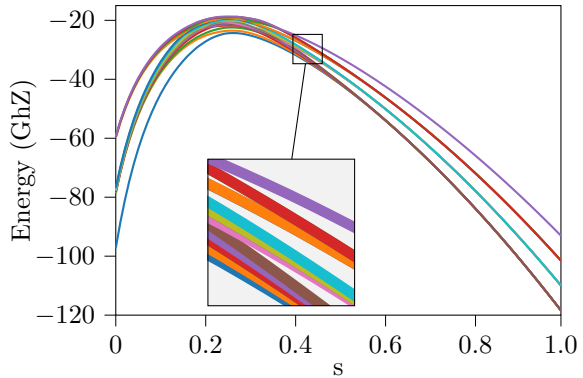}
\caption{Same as Figure~\ref{fig:linear} except using the schedule given in Figure~\ref{fig:sche} rather than a linear schedule given in Equation~(\ref{eq2}).   \label{fig:dwavesch}}
\end{figure}

Actual quantum hardware, however, rarely follows the relatively simple linear schedule considered above. We plot the data released by D-Wave about their annealing schedule in Figure~\ref{fig:sche} for the Advantage System 5.4. We observe immediately that this schedule differs substantially from a linear schedule. Using $H_p$ from Equation~(\ref{eq3}), the D-Wave data are then used in Equation~(\ref{eq1}) as the annealing schedule. The energy spectrum is shown in Figure~\ref{fig:dwavesch}. Although $H$ is the same, the new spectrum looks visibly different from that of the linear schedule in Figure~\ref{fig:linear}. The anti-crossings have shifted to a different range of $s$. Thus, unlike in the previous case, annealing back to $s\approx 0.8$ will most likely not work and a different value of $s\approx 0.4$ appears more appropriate.

In summary, the energy spectrum reveals that some paths that lead to the ground state and excited states have a low minimum gap between them, even for a very small graph coloring problem. Additionally, the value of $s$, where the minimum gaps come closest, depends strongly on the annealing schedule. Finally, the appropriate choice of the reverse distance combined with the quality of the input initial state can be decisive factors in our ability to reach the ground state.

Although the above example gives us a theoretical understanding, it must be remembered that this spectrum is the ideal case at zero temperature. Temperatures in an actual device, e.g., from D-Wave, can be around $20$ mK. Such low but finite temperatures can be expected to induce effects not accounted for in the theoretical zero-temperature case for annealing times considered in this work.

\begin{figure}
\begin{tikzpicture}

\definecolor{darkgray176}{RGB}{196,196,196}
\definecolor{steelblue31119180}{RGB}{91,80,190}
\definecolor{steelblue310}{RGB}{91,180,90}
\definecolor{steelblue3180}{RGB}{191,80,90}
\begin{axis}[
tick align=outside,
tick pos=left,
x grid style={darkgray176},
xmin=0, xmax=100,
xtick style={color=black},
y grid style={darkgray176},
ymin=0.0, ymax=1,
ytick style={color=black},
grid=both,
height = 4.3cm,
width = 7.5cm,
xlabel=Annealing time ($\mu s$),
ylabel= $s$,
ytick={0.0,0.25,0.50,0.75,1.0},
yticklabels={$0.00$,0.25,$0.50$,0.75,$1.00$},
xtick={0.0,9.5,18.6,27.6,36.7,45.7,54.8,63.8,72.9,81.9,90.9,100},
xticklabels ={0.0, ,18.6, ,36.7,  , ,  63.8, ,81.9, ,100},
]
\addplot [ultra thick, steelblue31119180]
table {%
0 1
9.54545454545454 0.9
18.5909090909091 0.8
27.6363636363636 0.7
36.6818181818182 0.6
45.7272727272727 0.5
54.7727272727273 0.5
63.8181818181818 0.6
72.8636363636364 0.7
81.9090909090909 0.8
90.9545454545455 0.9
100 1
};

\addplot [ultra thick, steelblue3180]
table {%
	0 1
	9.54545454545454 0.95
	18.5909090909091 0.9
	27.6363636363636 0.85
	36.6818181818182 0.8
	45.7272727272727 0.75
	54.7727272727273 0.75
	63.8181818181818 0.8
	72.8636363636364 0.85
	81.9090909090909 0.9
	90.9545454545455 0.95
	100 1
};

\addplot [ultra thick, steelblue310]
table {%
	0 1
	9.54545454545454 0.85
	18.5909090909091 0.7
	27.6363636363636 0.55
	36.6818181818182 0.4
	45.7272727272727 0.25
	54.7727272727273 0.25
	63.8181818181818 0.4
	72.8636363636364 0.55
	81.9090909090909 0.7
	90.9545454545455 0.85
	100 1
};

\end{axis}

\end{tikzpicture}
\caption{Reverse annealing paths for three different reverse distances: $0.25$, $0.50$, and $0.75$. A maximum of $12$ points are allowed to be defined, which are labeled on the x-axis. \label{fig:revsch}}
\end{figure}
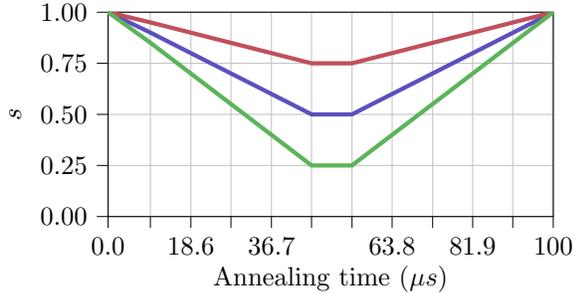

We fix the total annealing time for both forward and reverse annealing to 100 $\mu$s. The RA path defines the way in which the lowest reverse distance is reached and scaled back. We include a small pause in the schedule, which has been suggested to lead to better results~\cite{citerasc}. Three examples are shown in Figure~\ref{fig:revsch}. The D-Wave annealers take a maximum of $12$ inputs for the values of $s$. We use the same anneal path for all our experiments and from now on will only refer to the reverse distance, which is the lowest $s$ on the path.

\section{Results} \label{secres}

In this section, we thoroughly test RA for a variety of different cases for the graph coloring problem. We show the results obtained when feeding the bitstrings from forward annealing as the initial state of reverse annealing. We obtain $10^3$ forward annealing samples for each case in this section. If forward annealing finds one valid solution, this is taken as the initial classical state for the reverse annealing process. If forward annealing finds more than one valid solution, then one of the solutions is randomly chosen. This allows us to test if RA still works if a valid solution is used as an initial state. If forward annealing fails to find a valid solution, the sample with the lowest energy is chosen.

We divide the results into two sections. First, we study the quality of solutions obtained from reverse annealing as a function of the reverse distance. Second, we compare the scaling analysis of obtaining valid solutions using reverse and forward annealing for different problem sizes. We also use another device to illustrate that our proposal is device-independent. We always use the iterated version of reverse annealing where the total number of samples equals the number of iterated cycles.

\subsection{Reverse Distance}
\label{sec:rd}
\begin{figure}
\includegraphics[scale=0.55]{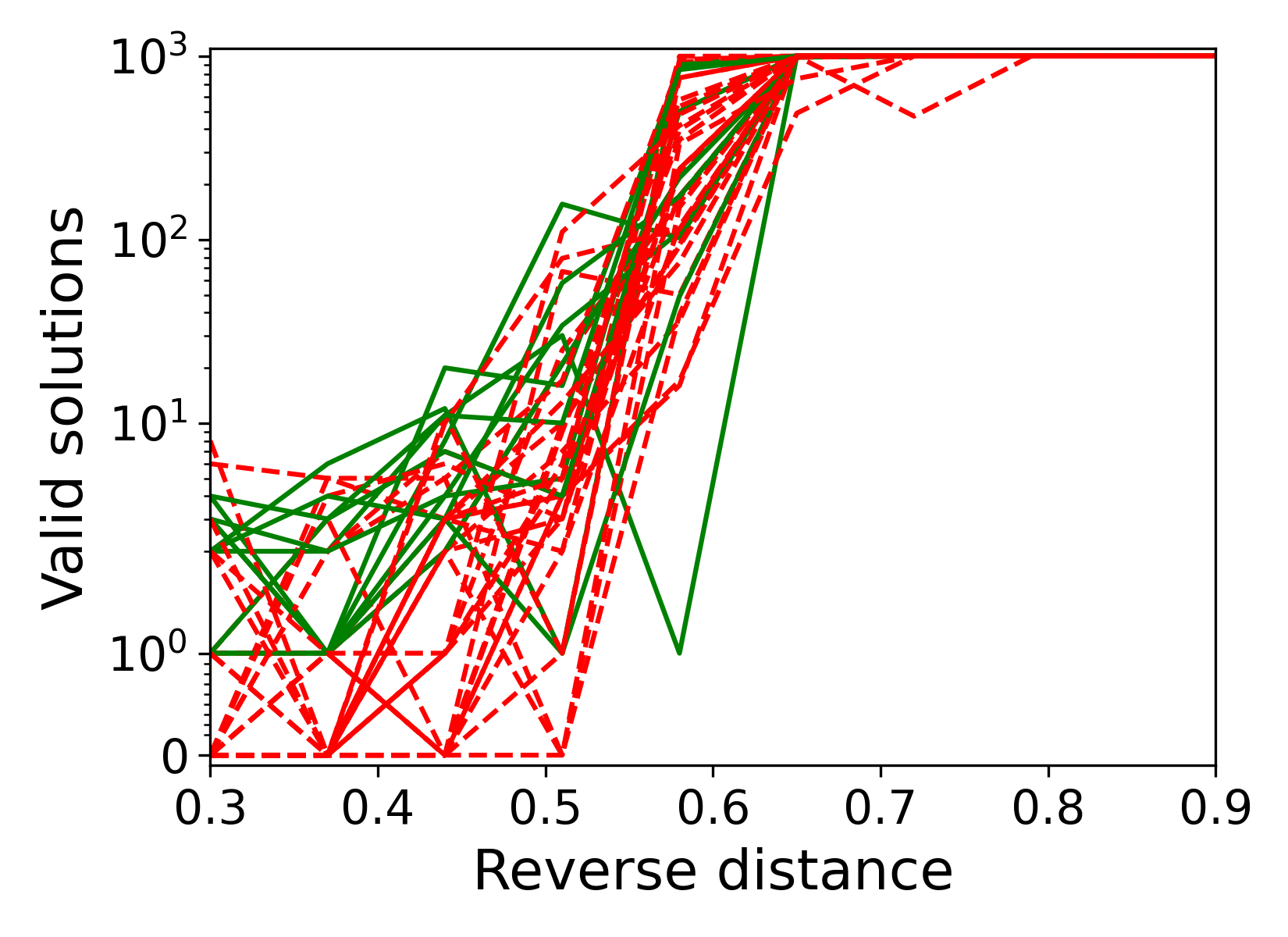}
\caption{{Total} number of valid solutions obtained as a function of the reverse distance. A valid solution from forward annealing was used as an initial state to the RA process. The graph represents a total of 41 out of 500 problems. Green (Red) lines represent that there was at least one valid solution at all values (at least one value) of the reverse distance. \label{figcasea}} 
\end{figure}

\begin{figure}
\includegraphics[scale=0.55]{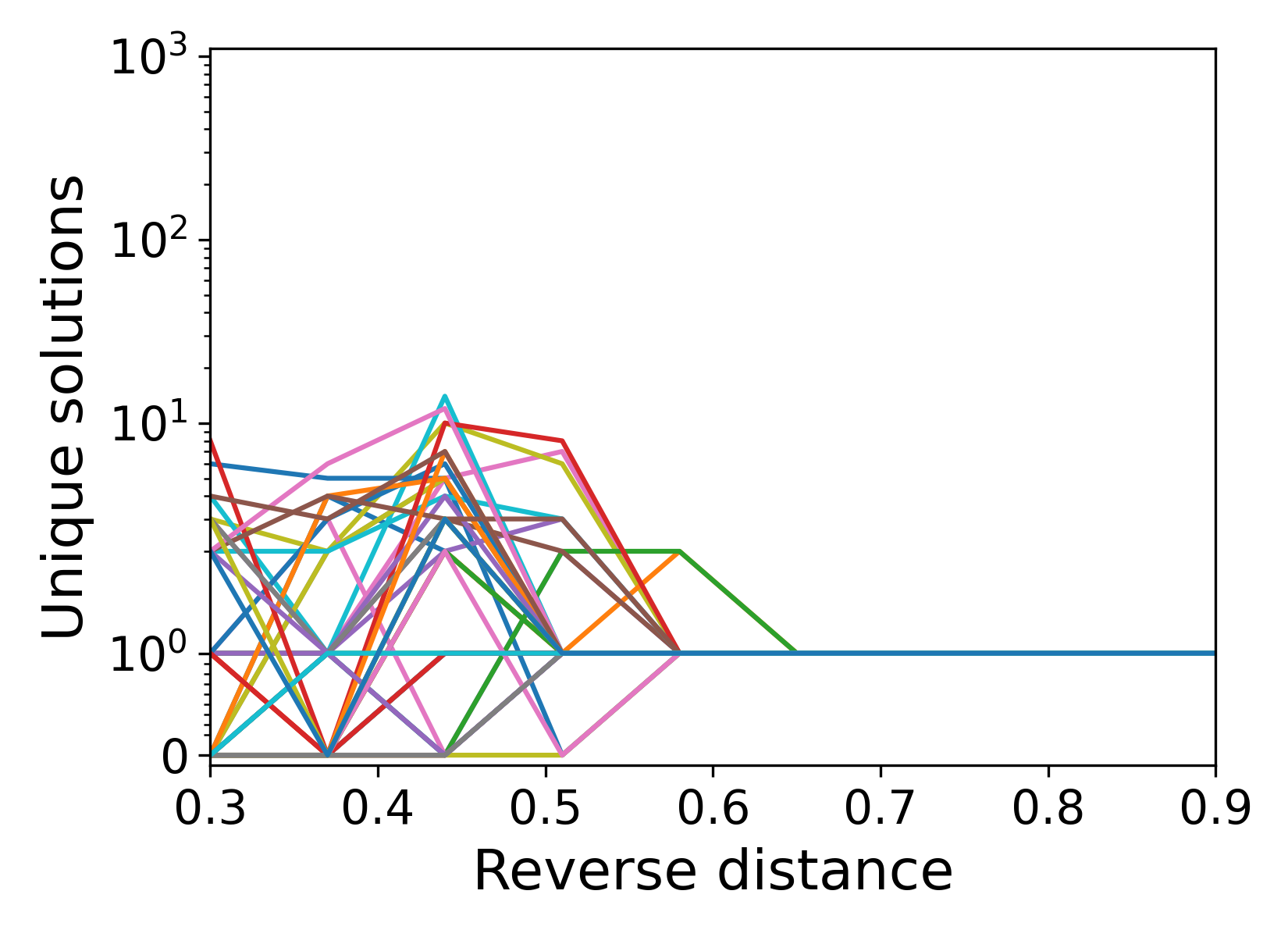}
\caption{Same as Figure~\ref{figcasea} except the y-axis shows total number of unique valid solutions. Each problem is given by a random color. \label{figcased}} 
\end{figure}

\begin{figure}
\includegraphics[scale=0.55]{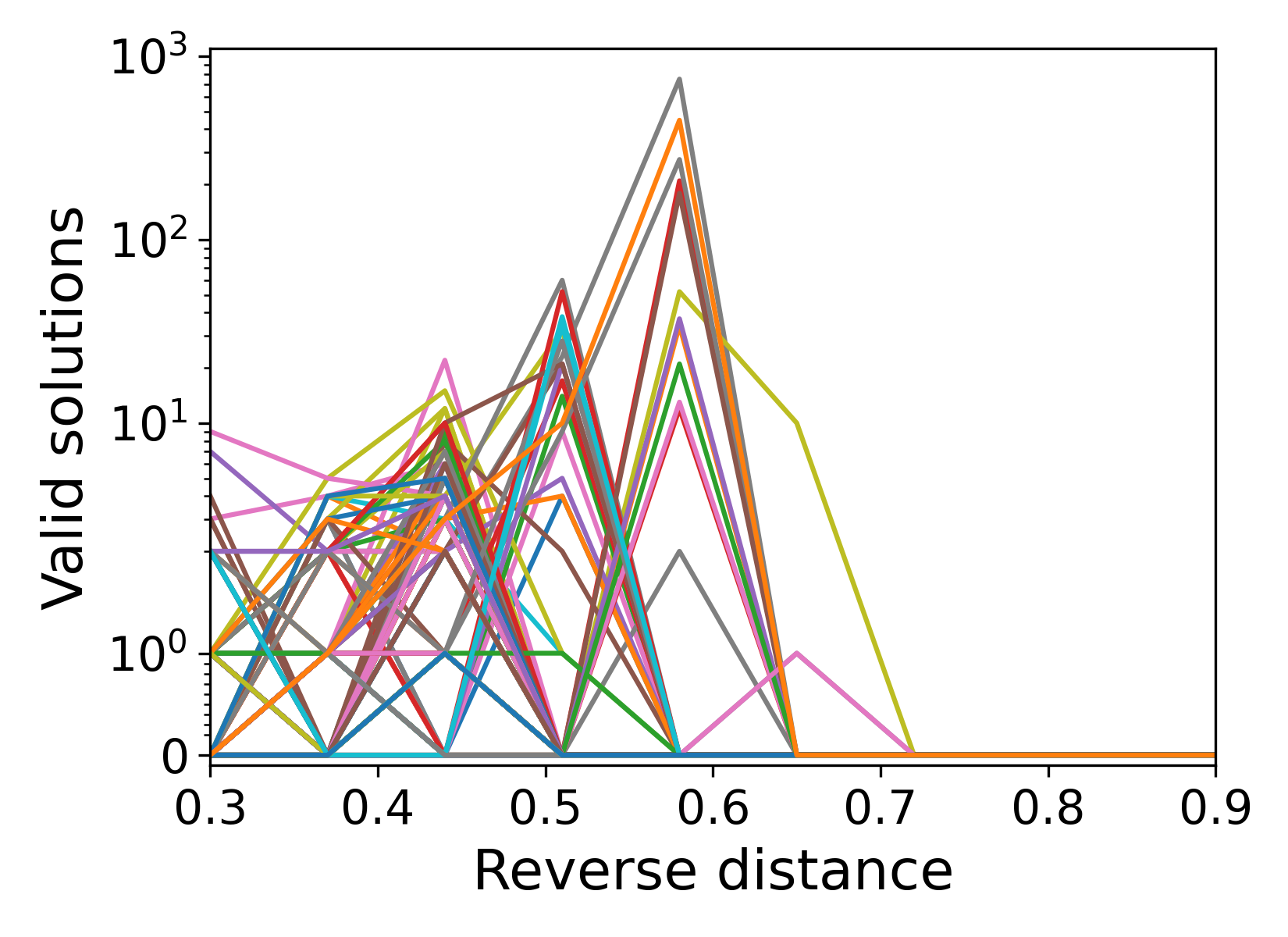}
\caption{Total number of valid solutions obtained as a function of the reverse distance. An invalid solution with the lowest energy from forward annealing was used as an initial state to the RA process. The graph represents a total of 262 out of 500 problems.  \label{figcase1}}
\end{figure}

\begin{figure}
\includegraphics[scale=0.55]{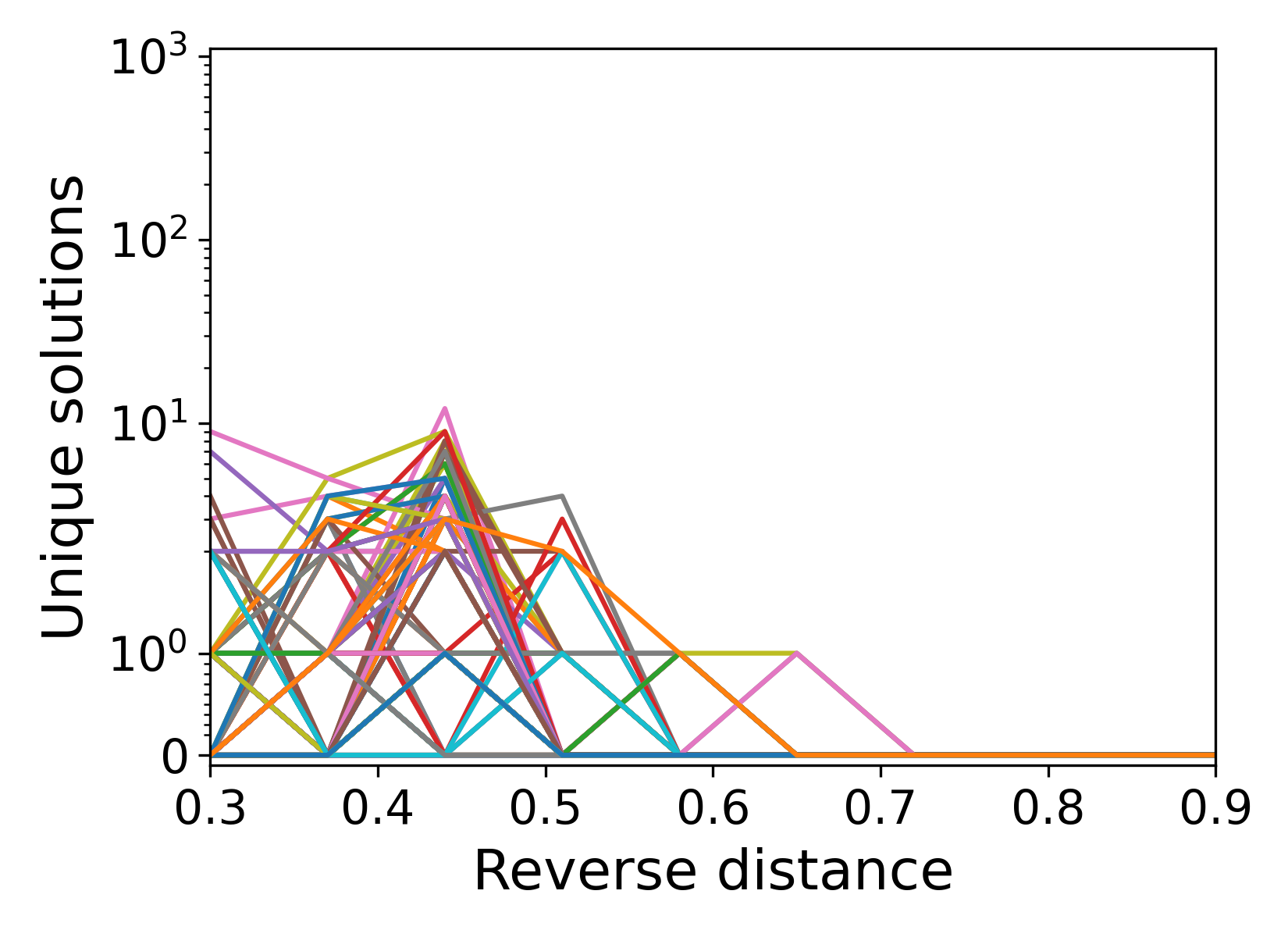}
\caption{Same as Figure~\ref{figcase1} except that the y-axis shows the total number of unique valid solutions. \label{figcase2}}
\end{figure}

{In order to demonstrate our idea, we create $500$ random graphs generated according to the Erdős--Rényi model~\cite{SciPyProceedings_11}, and the number of vertices is set to $15$. The value of $k$ is obtained from the greedy algorithm. To test assisted reverse annealing, the reverse distance is decreased from $0.93$ to $0.30$ in a total of $10$ equally spaced steps. For each value of the reverse distance, we obtain $10^3$ samples. We analyze the total number of valid solutions as well as the total number of unique valid solutions. Since the initial state of the RA process can itself either be a valid or an invalid solution, depending on whether the forward annealing process was able to find at least one valid solution out of $10^3$ samples, we then have four cases:}
\begin{enumerate}
\item Valid RA solutions; valid solution initial state. See Figure~\ref{figcasea}.
\item Unique RA solutions; valid solution initial state. See Figure~\ref{figcased}.
\item Valid RA solutions; invalid solution initial state. See Figure~\ref{figcase1}.
\item Unique RA solutions; invalid solution initial state. See Figure~\ref{figcase2}.
\end{enumerate}

\subsubsection{{Case A}} 
Shown in Figure~\ref{figcasea} are the total numbers of valid solutions obtained at different values of the reverse distance. Plotted are only $41$ out of $500$ problems because forward annealing was able to find a valid solution only in these cases. Since we start from a valid solution obtained from forward annealing, we observe that almost all obtained samples are valid solutions at larger values of the reverse distance. This behavior is expected between a reverse distance range of $0.75$ and $1.0$. One explanation arises from the fact that the function $B(s)$ changes only negligibly between $0.75$ and $1.0$ in the annealing schedule of the D-Wave annealer shown in Figure~\ref{fig:sche}. Due to this, the $H$ in Equation~(\ref{eq1}) has $H_p$ as the dominant component. Therefore, upon reverse annealing, no major contribution is expected from the mixing term $H_i$ and the annealer is expected to stay in the provided initial state.

The number of valid solutions decreases drastically as the reverse distance is decreased further below $0.75$. The mixing term $H_i$ also becomes relevant and the annealer is no longer able to sample only the initial state it was provided, leading to a decrease in the number of valid solutions.

We also observe that there are cases in which at least one valid solution was obtained for each value of the reverse distance, shown in green. However, in most cases, not all values of the reverse distance yield at least one valid solution, as shown by the red (dotted) lines. The fact that forward annealing was able to find a valid solution in only $8.2\%$ of the total cases is an indication that the problem size is appropriate to test whether reverse annealing can obtain valid solutions for the rest of the cases.

\subsubsection{Case B}
Deterministic greedy algorithms find one of all possible valid solutions if the solution space is degenerate. A quantum annealer may offer the advantage that one may obtain different unique solutions by no additional effort than sampling more often than once. We study this case by filtering out the number of unique valid solutions obtained from Case A and plot the results in Figure~\ref{figcased}.

We observe that at large reverse distances, all problems converge to the value of one unique solution. The unique solution is clearly separate for each problem. This solution is the same one that we use as the initial state, which was in turn obtained from forward annealing. Thus, from Figures~\ref{figcasea} and~\ref{figcased}, we discern that we exclusively sample the initial state input to the RA process for larger annealing distances.

{The results take an exciting turn at lower reverse distances. We observe a novel pattern where unique solutions start to emerge below a reverse distance of $0.60$, reach a peak at $0.44$, and then decline again at $0.30$. This result is interesting because we showed previously in Figure~\ref{fig:dwavesch} that the anti-crossings appear close to $0.4$, albeit for a smaller problem size. Thus, it is possible to sample more unique solutions simply by choosing an appropriate reverse distance. }

\subsubsection{Case C}
\label{casec}
The most interesting test for reverse annealing is when forward annealing finds no valid solutions. This was the case for $459$ out of $500$ problems. We take the lowest energy bitstring as the initial state and plot the total number of valid solutions obtained as a function of the reverse distance. The results are shown in Figure~\ref{figcase1}.

In contrast to Case A, we observe no valid solutions for all problems at larger reverse distances. This behavior is expected since, at large values, the annealer outputs the same invalid bitstring input as the initial state. Valid solutions start to appear for reverse distances less than $0.7$, and the highest total of these valid solutions tends to decrease as the reverse distance decreases.

At smaller reverse distances, we were able to obtain at least one valid solution for $262$ out of $500$ problems. Figure~\ref{figcase1} only shows these $262$ problems; we omit those for which reverse annealing also found no valid solution. The peak number of total solutions obtained differs for each problem across the reverse distance. From Case C, we conclude that reverse annealing is able to find valid solutions in about $57\%$ of problems where forward annealing does not.

\subsubsection{Case D}
We plot the unique solutions obtained in Case C in Figure~\ref{figcase2}. We observe that the number of unique solutions first increases as the reverse distance is decreased, peaks at $0.44$, and then decreases when further decreasing the reverse distance. A majority of the problems had the maximum number of unique solutions at a reverse distance of $0.44$. In contrast to Case B, there are no valid solutions at larger reverse distances; hence, all lines converge to the value $0$.

\subsection{Scaling Analysis}

In the previous section, we observed that reverse annealing is able to find valid solutions where forward annealing fails. In this section, we perform an analysis across different problem sizes. In addition to the data for the $500$ graphs with $15$ vertices shown in the previous section, we generate $100$ new random graphs, each with $7, 10, 17$, and $20$ vertices. 

Since all the graphs are generated randomly, it is not necessary that all of them having the same node size can be colored using the same number of colors. The number of qubits required is also dependent on the number of colors needed. Hence, there will be a distribution of different number of qubits used across the $100$ (or $500$) random graphs of a given node size. Therefore, we chose the node sizes such that we sufficiently cover all relevant qubit numbers for our scaling analysis. The software from D-Wave automatically creates chains of physical qubits to form logical qubits to accommodate larger problems. We fix the reverse distance to $0.44$ for all problems. This choice is inspired from the results of Case B from Section~\ref{sec:rd}. 

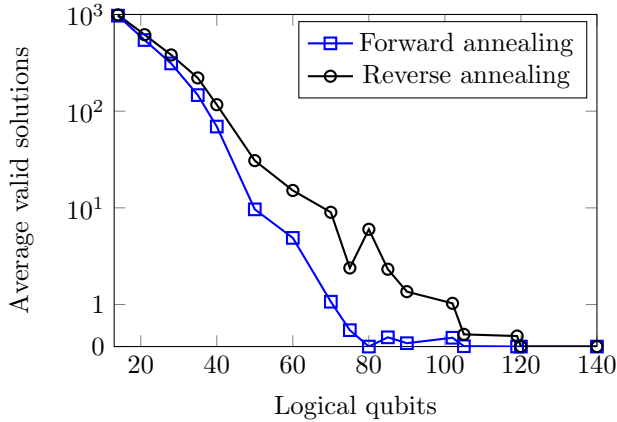
\begin{figure}
\begin{tikzpicture}

\definecolor{darkgray176}{RGB}{176,176,176}
 \def\basis{1}
\pgfplotsset
{
	y coord trafo/.code={\pgfmathparse{symlog(#1,\basis)}\pgfmathresult},
	y coord inv trafo/.code={\pgfmathparse{symexp(#1,\basis)}\pgfmathresult},
	yticklabel style={/pgf/number format/.cd,int detect,precision=2},
}
	\begin{axis}[
ylabel=Average valid solutions,
height = 6cm,
width = 8cm,
scaled ticks = base 10:0,
ymin=0, ymax = 1000,   
xmin = 13, xmax = 140,
xlabel=Logical qubits,
ytick = {0,1,10,100,1000},
yticklabels= {$0$,$1$,$10^1$,$10^2$,$10^3$},
		]
\addplot[mark=square, blue, mark size = 2.2pt, thick] table {
X Y
14 971
21 544.23
28 311.55
35 146.6
40 69.3
50 9.65
60 4.89
70 1.07
75 0.39
80 0
85 0.22
90 0.08
102 0.21
105 0.009
119 0 
120 0
140 0
160 0
180 0 
};
\addlegendentry{Forward annealing} 

\addplot[mark=o, black, thick, mark size = 2.1pt] table {
X Y
14 990
21 617.7
28 381.42
35 220.6
40 116.67
50 30.79
60 15.13
70 9
75 2.39
80 6
85 2.32
90 1.36
102 1.03
105 0.29
119 0.25
120 0
140 0 
160 0
180 0
};
\addlegendentry{Reverse annealing} 
\end{axis}
\end{tikzpicture}
\caption{Average valid solutions obtained when using forward and reverse annealing with a reverse distance of $0.44$ across 900 randomly generated problems. \label{figmain1}}
\end{figure}

We calculate the average number of valid solutions obtained for all problems that needed the same logical qubit count. The data are plotted in Figure~\ref{figmain1}. The number of problems per qubit number differs between different qubit numbers. 

We observe that the frequency of success vanishes exponentially in problem size for both forward and reverse annealing. For small problem sizes, given by the corresponding small number of logical qubits required (i.e., <40), we observe that both forward and reverse annealing work almost equally well. In the next region between $40$ and $80$ qubits, the frequency of finding valid solutions using forward annealing drops further exponentially to below $1$. Beyond $80$ qubits, forward annealing completely fails to find valid solutions on average. In contrast, reverse annealing is able to find more valid solutions on average in all cases up to $80$ qubits. The most interesting cases happen beyond $80$ qubits where reverse annealing is able to find at least one valid solution on average in contrast to forward annealing. The average number of valid solutions for reverse annealing drop further until $119$ qubits.

At $120$ qubits, both forward and reverse annealing yield no valid solution for any of the problems. These problems have our largest node size of $20$. Our obtained data {(not shown)} for $160$ and $180$ qubits yield also zero valid solutions for all tested problems.

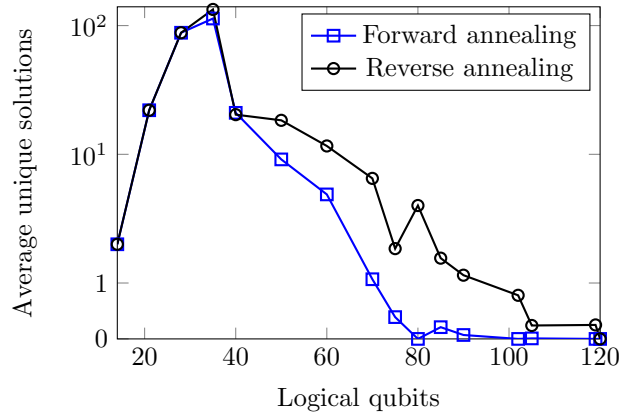
\begin{figure}
\begin{tikzpicture}

\definecolor{darkgray176}{RGB}{176,176,176}
 \def\basis{1}
\pgfplotsset
{
	y coord trafo/.code={\pgfmathparse{symlog(#1,\basis)}\pgfmathresult},
	y coord inv trafo/.code={\pgfmathparse{symexp(#1,\basis)}\pgfmathresult},
	yticklabel style={/pgf/number format/.cd,int detect,precision=2},
}
	\begin{axis}[
ylabel=Average unique solutions,
height = 6cm,
width = 8cm,
scaled ticks = base 10:0,
ymin=0, ymax = 140,   
xmin = 14, xmax = 120,
xlabel=Logical qubits,
ytick = {0,1,10,100,1000},
yticklabels= {$0$,$1$,$10^1$,$10^2$,$10^3$},
		]
\addplot[mark=square, blue, mark size = 2.2pt, thick] table {
X Y
14 2
21 22
28 88
35 113.5
40 21
50 9.13
60 4.88
70 1.07
75 0.39
80 0
85 0.21
90 0.07
102 0
105 0.009
119 0
120 0
140 0
160 0
180 0
};
\addlegendentry{Forward annealing} 

\addplot[mark=o, black, thick, mark size = 2.1pt] table {
X Y
14 2
21 22
28 88
35 133.5
40 20.33
50 18.37
60 11.61
70 6.5
75 1.85
80 4
85 1.56
90 1.15
102 0.78
105 0.24
119 0.25
120 0
140 0
160 0
180 0
};
\addlegendentry{Reverse annealing} 
\end{axis}
\end{tikzpicture}
\caption{Same as Figure~\ref{figmain1} except the y-axis shows average of unique valid solutions. \label{fig22}}
\end{figure}

We now move to analyzing average unique solutions obtained. The data are shown in Figure~\ref{fig22}. We observe that, at first, the number of average unique solutions increases, reaches a peak, and then decreases. This is explained by the fact that for small problem sizes, there does not exist a large number of degenerate solutions and the quantum annealer is able to find all possible unique solutions using both forward and reverse annealing. For large problem sizes, neither forward nor reverse annealing are able to find all unique solutions. Increasing the problem size further then shows an exponential decline in the average number of unique solutions found. Here, the plot exhibits the same features as the plot in Figure~\ref{figmain1}. In summary, reverse annealing is able to find more unique solutions on average than forward annealing for the problem sizes we tested.

In conclusion, from Figures~\ref{figmain1} and~\ref{fig22}, we infer that there exists a range of problem sizes for which one can expect forward annealing to find no valid solutions, but feeding the results to reverse annealing yields valid solutions. It is in this sense that reverse annealing extends the computational reach of quantum annealers for the graph coloring problem.

\subsection{Random Initial State}

Since reverse annealing accepts any bitstring as initial state, one may question the need for first performing forward annealing as outlined in our algorithm. We now compare our idea of using the lowest-energy bitstring against using a random bitstring as an initial state. Additionally, we repeat our experiments on a different device to confirm that the algorithm is not device dependent.

\begin{figure}
\begin{tikzpicture}

\definecolor{darkgray176}{RGB}{176,176,176}
\definecolor{darkgreen176}{RGB}{110,220,110}
	\begin{axis}[
ylabel=Average valid solutions,
height = 6cm,
width = 8cm,
ymin=0, ymax = 2,   
xmin = 0.3, xmax = 0.93,
xlabel=Reverse distance,
		]

\addplot[mark=o, red, thick, mark size = 2.1pt] table {
X Y
0.3 0.046
0.37 0.065
0.44 0.271
0.51 0.0
0.58 0
0.65 0
0.72 0
0.79 0
0.86 0
0.93 0
};
\addlegendentry{\small Random bitstring}


\addplot[mark=triangle, black, thick, mark size = 2.1pt] table {
X Y
0.3 0.164
0.37 0.141
0.44 1.252
0.51 0.164
0.58 0
0.65 0
0.72 0
0.79 0
0.86 0
0.93 0
};
\addlegendentry{\small Annealer \#2} 

\addplot[mark=square, darkgreen176, mark size = 2.2pt, thick] table {
X Y
0.3 0.47
0.37 0.57
0.44 1.85
0.51 1.55
0.58 0.65
0.65 0.05
0.72 0
0.79 0
0.86 0
0.93 0
};
\addlegendentry{\small Best bitstring}

\end{axis}
\end{tikzpicture}
\caption{Average valid solutions obtained for 262 problems as a function of the reverse distance. Squares represent the case of using the lowest-energy bitstring from forward annealing as the initial state. Triangles represent the use of another device. Circles represent the case where the initial state was a random bitstring.\label{lastfig}}
\end{figure}
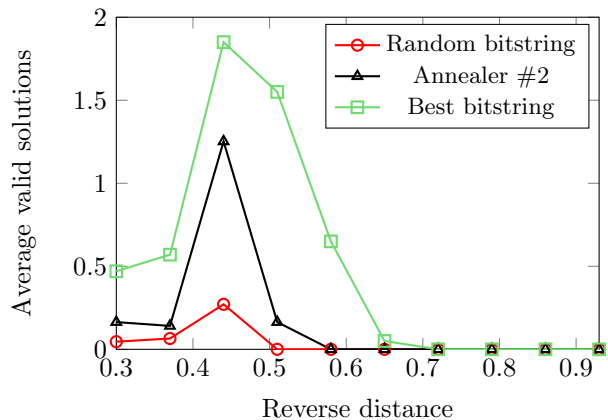

For each of the $262$ problems from Section \ref{casec} Case C, we repeat the experiment only for those values of the reverse distance for which reverse annealing was able to find at least one valid solution with a slight change that the initial state is now a random bitstring. This allows us to save significant computing time on the real hardware. The data are shown in Figure~\ref{lastfig}. We plot the average number of valid solutions obtained for cases when using a random initial state in red circles. Next, we plot the average number of valid solutions when using the lowest energy (invalid solution) bitstring from forward annealing in green squares. Next, we repeat the green squares case with $3000$ samples on a different annealer, Advantage System 6.2.

We observe that in a small number of cases, starting from a random initial state, the device is still able to find a valid solution for small reverse distances. In contrast, starting from the lowest energy bitstring from forward annealing yields significantly higher averages over a wider region of reverse distances. Additionally, we observe that a different annealer is able to reproduce some results but does not perform better despite a higher number of samples.

The lower performance of a different device can be explained as follows. In solving a given problem, several device parameters are involved, e.g., the annealing time, connectivities between the qubits, embedding, as well as chain strength. One set of such parameters that yields higher frequency of valid solutions on one device need not necessarily yield similar results on another device. We used the same parameters for both devices; thus, there may exist other parameters for the second device that yield a higher frequency of valid solutions. However, finding these parameters is beyond the scope of the current work.

In summary, we learned that there is a significant benefit in first performing forward annealing and choosing the lowest energy bitstring compared to using a bitstring created at random. 

\section{Conclusions}\label{secconc}

Finding appropriate initial states for reverse annealing to improve is, in general, an important and difficult task.
We proposed a heuristic where the output of forward annealing was used as input for reverse annealing. We found that this setup is particularly useful for the graph coloring problem. We demonstrated our approach by solving randomly generated graph coloring problems. We found that reverse annealing is able to find valid solutions using this technique in about $57\%$ of the cases where forward annealing fails. Thus, one is able to solve many more problems outside the reach of conventional forward annealing with the same quantum annealer.

We performed a scaling analysis where we increased the problem size and used both forward and the new forward-assisted reverse annealing to solve the same problem. Our results showed an exponential decay in the probability to sample valid solutions for both approaches. However, the new approach decayed slower and, thus, is able to extend the computational reach of the overall heuristic of quantum annealing. The new heuristic proposed in this work is general and can be used for any problem that uses forward annealing.

While we showed the working of the new approach, our results do not reflect best effort, which would include finding the most optimal annealing times, reverse distances, chain strengths, embeddings, annealing schedules, and other relevant parameters. These should help obtain valid solutions in those cases where reverse annealing was also not able to find a valid solution. This can be explored in future works.

\section*{Data availability}
The data has been made publically available at \url{https://doi.org/10.25716/gude.0xjr-gh7d} \cite{rawdata}.

\section*{Acknowledgements}
The author gratefully acknowledges the Jülich Supercomputing Centre (https://www.fz-juelich.de/ias/jsc) for funding this project by providing computing time on the D-Wave Advantage System JUPSI through the Jülich UNified Infrastructure for Quantum computing (JUNIQ). Special thanks to Fengping Jin, Jens-Bastian Eppler, and Lucas Joshua Menger for their feedback.

\end{document}